\documentclass[prd,aps,preprint,showpacs,nofootinbib,superscriptaddress,tightenlines]{revtex4-1}
\usepackage{epsfig}
\usepackage{amssymb}
\usepackage{amsmath}
\usepackage{graphicx}
\usepackage{psfrag}

\begin{document}

\title{$k_t$-factorization for Hard Processes in Nuclei}
\author{Fabio Dominguez}
\affiliation{Department of Physics, Columbia University, New York,
NY, 10027}
\author{Bo-Wen Xiao}
\affiliation{Nuclear Science Division, Lawrence Berkeley National
Laboratory, Berkeley, CA 94720}
\author{Feng Yuan}
\affiliation{Nuclear Science Division, Lawrence Berkeley National
Laboratory, Berkeley, CA 94720} \affiliation{RIKEN BNL Research
Center, Building 510A, Brookhaven National Laboratory, Upton, NY
11973}
\begin{abstract}
Two widely proposed $k_t$-dependent gluon distributions in
the small-$x$ saturation regime are investigated using two particle
back-to-back correlations in high energy scattering processes.
The Weizs\"{a}cker-Williams gluon distribution, interpreted 
as the number density of gluon inside nucleus, is studied 
in the quark-antiquark jet correlation in deep inelastic 
scattering. On the other hand, the
unintegrated gluon distribution, defined as the Fourier
transform of the color-dipole cross section, is probed
in the direct photon-jet correlation in $pA$ collisions.
Dijet-correlation in $pA$ collisions depends on both gluon
distributions through combination and convolution in the large
$N_c$ limit. 
\end{abstract}

\maketitle

%\preprint{LBNL-xxx}\preprint{RBRC-xxx}

%\email{bxiao@lbl.gov}

%\email{fyuan@lbl.gov}

%\section{Introduction}

As the foundation of high energy hadronic physics, QCD
factorization enables us to separate the short distance
perturbative physics from the long distance non-perturbative
effects. Its prediction power relies on the universality of
the parton distributions among different processes.
Recent studies~\cite{BoeVog03,Bomhof:2006dp,qvy-short,Collins:2007nk,
Vogelsang:2007jk,Rogers:2010dm,Xiao:2010sp} 
have shown that the naive $k_t$-factorization
is violated in two-particle production in hadron-hadron
collisions. Nevertheless, in this letter, we establish an effective
factorization in hard processes in nuclei scattered by
a dilute probe by modifying the parton distributions at
small-$x$. Although the parton distributions are no longer
universal, they can be constructed from several universal
individual building blocks.

The saturation phenomena in high energy hadronic interactions
have attracted great attention in recent years.
It has long been recognized that the gluon dynamics in QCD
at small-$x$ is responsible for these phenomena~\cite{Gribov:1984tu,Mueller:1985wy,McLerran:1993ni,{Iancu:2003xm}}.
An effective theory, the color glass condensate (CGC), was
proposed to systematically study this physics~\cite{Iancu:2003xm}.
On the experimental side, there exist strong indications of the saturation
from the structure function measurements in deep inelastic
scattering (DIS) at HERA and the shadowing effects found
in inclusive hadron production in $dA$ collisions at
RHIC~\cite{Iancu:2003xm}. Ongoing and future experiments
from both RHIC and LHC, and in particular, the planned Electron-Ion
Collider~\cite{Deshpande:2005wd}, shall provide further information
on this.

An important aspect of the small-$x$ gluon distribution function
in nucleons and nuclei is the resummation of the multiple
interactions of the hadronic probe with the target, because the
gluon density is so high and these interactions have to be taken
into account. In order to study the resummation effects, a
transverse momentum dependence is introduced. They are referred to as
the $k_t$-dependent gluon distributions, also called unintegrated
gluon distribution (UGD) functions. Two different forms of the
UGDs have been used in the literature. The first gluon
distribution, also known as the Weizs\"{a}cker-Williams (WW)
gluon distribution, measures the number density of gluons in the
CGC formalism~\cite{McLerran:1993ni}, whereas the second one
defined as the Fourier transform of the color dipole cross
section, appears in the calculations for, e.g., single inclusive
particle production in $pA$ collisions~\cite{Iancu:2003xm}.
However, it has been argued that they can not be distinguished
especially in the small $k_{\perp}$ region though they differ
dramatically, and both of them are often
used~\cite{Kharzeev:2003wz}.

In this paper, we study two particle correlations in various high
energy scattering processes as probes to these UGDs. There have
been intensive investigations of these processes in the last few
years~\cite{BoeVog03,Bomhof:2006dp,qvy-short,Collins:2007nk,
Vogelsang:2007jk,Rogers:2010dm,Xiao:2010sp}. Taking the quark distribution
as an example, it was shown in
Ref.~\cite{Xiao:2010sp} that, in the large nuclear number limit,
an effective factorization can be achieved with modified parton
distributions of nucleus in $pA$ and $\gamma^*A$ scattering
processes where the multiple interaction effects can be resummed
in the small-$x$ formalism. Following this argument, we focus on
the processes with a dilute system scattering on a dense target,
\begin{equation}
B+A\to H_1(k_1)+H_2(k_2)+X \ ,
\end{equation}
where $A$ represents the dense target (such as a nucleus), $B$ stands
for the dilute projectile (such as nucleon or photon),
$H_1$ and $H_2$ are the two final state particles with momenta
$k_1$ and $k_2$, respectively. We are interested in
the kinematic region where the transverse momentum imbalance between
them is much smaller than the individual momentum:
$q_{\perp }=|\vec{k}_{1\perp }+\vec{k}_{2\perp }|\ll P_\perp$
where $\vec{P}_\perp$ is defined as
$(\vec{k}_{1\perp }-\vec{k}_{2\perp })/2$.
This is referred as the back-to-back correlation limit (the
correlation limit) in the following discussions.
An important advantage of taking this limit
is that we can apply the power counting method to
obtain the leading order contribution of
$q_{\perp }/P_{\perp }$ where the differential cross section
directly depends on the UGDs of the nuclei.
For example, the quark-antiquark jet correlation in deep
inelastic scattering (DIS) directly probes the first type of the UGD, whereas
the direct photon-quark jet correlation in $pA$ collisions
probes the second type of UGD.
The dijet (di-hadron) correlations in $pA$ collisions can probe
both gluon distributions, though the connection is more
complicated.

Let us first discuss the conventional gluon distribution, generalized to
include the transverse momentum dependence~\cite{Collins:1981uw,{Ji:2005nu}},
\begin{eqnarray}
xG^{(1)}(x,k_{\perp })&=&\int \frac{d\xi ^{-}d^2\xi _{\perp }}{(2\pi )^{3}P^{+}}%
e^{ixP^{+}\xi ^{-}-ik_{\perp }\cdot \xi _{\perp }}\nonumber\\
&&\times \langle P|F^{+i}(\xi
^{-},\xi _{\perp })\mathcal{L}_{\xi }^{\dagger }\mathcal{L}%
_{0}F^{+i}(0)|P\rangle \ ,\label{g1}
\end{eqnarray}%
where $F^{\mu \nu }$ is the gauge field strength tensor $F_a^{\mu
\nu }=\partial ^{\mu }A_a^{\nu }-\partial ^{\nu }A_a^{\mu
}-gf_{abc}A_b^\mu A_c^\nu$ with $f_{abc}$ the antisymmetric structure
constants for $SU(3)$, and $\mathcal{L}_{\xi }={\cal
P}\exp\{-ig\int_{\xi ^{-}}^{\infty }d\zeta ^{-}A^{+}(\zeta ,\xi
_{\perp })\}{\cal P}\exp\{-ig\int_{\xi _{\perp }}^{\infty }d\zeta
_{\perp }\cdot A_{\perp }(\zeta ^{-}=\infty ,\zeta _{\perp })\}$
is the gauge link in the adjoint representation $A^{\mu
}=A_{a}^{\mu }t_{a}$ with $t_{a}=-if_{abc}$. It contains a
transverse gauge link at spatial infinity which is important to
make the definition gauge invariant~\cite{Belitsky:2002sm}. 
These gauge links have to be made non-light-like to
regulate the light-cone singularities when gluon radiation contributions
are taken into account~\cite{Collins:1981uw}. This does not affect the following analysis, because there is no light-cone singularity in the calculation. By
choosing the light-cone gauge with certain boundary condition for
the gauge potential (for example, in the above definition,
$A_{\perp }(\zeta ^{-}=\infty)=0$), we can drop out the gauge link
contribution, and find that this gluon distribution has the number
density interpretation. Then, it can be calculated from the wave
functions or the WW field of the nucleus
target~\cite{McLerran:1993ni,{Kovchegov:1998bi}}. At small-$x$ for
a large nucleus, it was found~\cite{McLerran:1993ni}
\begin{equation}
xG^{(1)}(x,k_{\perp })=\frac{S_{\perp }}{\pi ^{2}\alpha _{s}}\frac{%
N_{c}^{2}-1}{N_{c}}\int \frac{d^{2}r_{\perp }}{(2\pi )^{2}}\frac{%
e^{-ik_{\perp }\cdot r_{\perp }}}{r_{\perp }^{2}}\left( 1-e^{-\frac{r_{\perp
}^{2}Q_{s}^{2}}{4}}\right) \ ,
\end{equation}%
where $N_c=3$ is the number of colors and
$Q_{s}$ is the gluon saturation scale~\cite{Iancu:2003xm}.
We have cross checked this result by directly calculating the
gluon distribution function in Eq.~(\ref{g1}) following the similar
calculation for the quark in Ref.~\cite{Belitsky:2002sm, Brodsky:2002ue}.

\begin{figure}[tbp]
\begin{center}
\includegraphics[width=12cm]{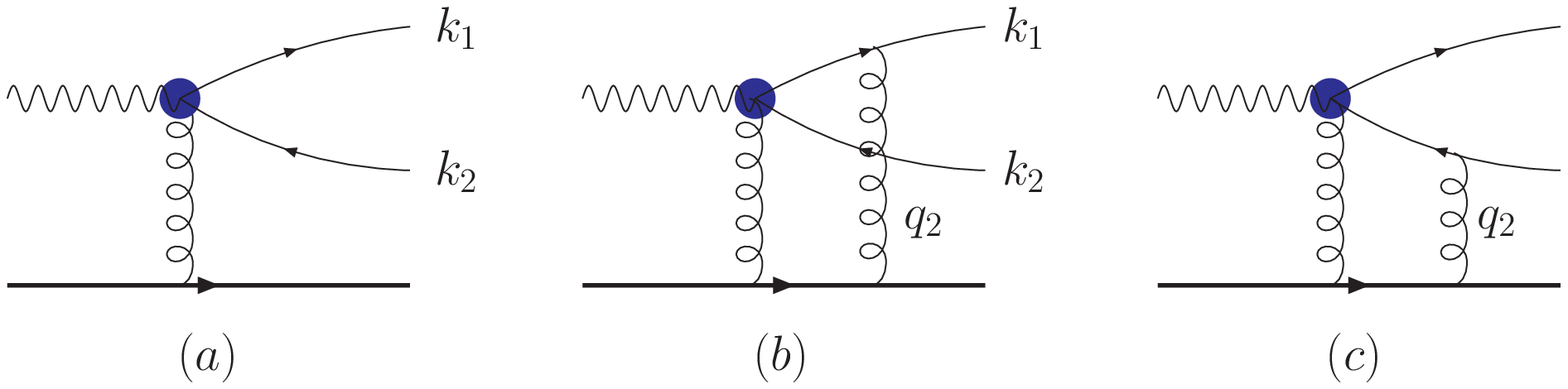} \\ %\hfill
\includegraphics[width=12cm]{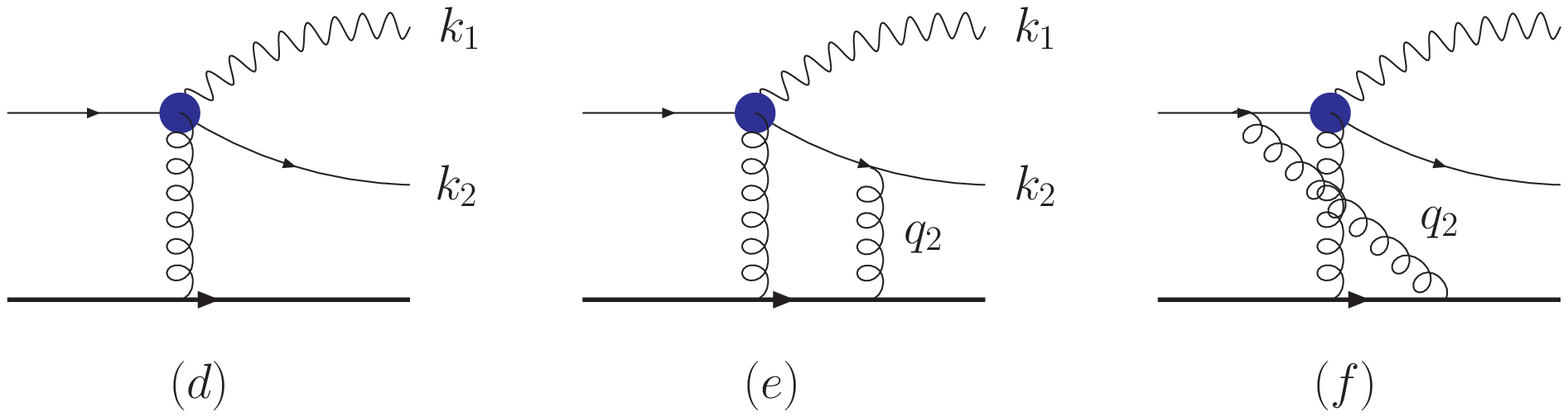}%\hfill
\end{center}
\caption[*]{Typical Feynman diagrams contributing to the
quark-antiquark jet correlation in deep inelastic scattering (a,b,c)
and photon-jet correlation in $pA$ collisions (d,e,f):
(a,d) leading order, where the bubble represents the gluon attachments to both
quark lines; (b,c,e,f) two-gluon exchange diagrams. }
\label{fig1}
\end{figure}

Despite the nice physical interpretation, it has been argued that
the gluon distribution in Eq.~(\ref{g1}) is not directly related
to physical observables in the CGC formalism. However, we will
show that $xG^{(1)}$ can be directly probed through the
quark-antiquark jet correlation in DIS,
\begin{equation}
\gamma_T^{\ast }A\rightarrow q(k_{1})+\bar{q}(k_{2})+X\ .\label{dis}
\end{equation}
The calculations are performed for $Q^2$ in the same order of
$P_\perp^2$ and the quarks are massless. Extension to the real
photon scattering and/or massive quarks in the final state is
straightforward. We will show the results for the transversely
polarized photon, and that for the longitudinal one follows
accordingly. We plot the typical Feynman diagram for
the process of (\ref{dis}) in Fig.~1, where the bubble in the
partonic part represents the hard interaction vertex including
gluon attachments to both quark and antiquark lines. Fig.~1(a) is
the leading Born diagram whose contributions can be associated with the
hard partonic cross section times the gluon distribution from
Eq.~(\ref{g1})~\cite{qvy-short}. In high energy scattering with
the nucleus target, additional gluon attachments are important and
we have to resum these contributions in the large nuclear
number limit. Figs.~1(b,c) represent the diagrams contributing at
two-gluon exchange order, where the second gluon can attach to
either the quark line or the antiquark line. By applying the power
counting method in the correlation limit ($q_\perp\ll P_\perp$),
we can simplify the scattering amplitudes with the Eikonal
approximation~\cite{qvy-short}. For example, Fig.~1(b) can be
reduced to: $g/(-q_{2}^{+}+i\epsilon )T^{b}\Gamma^{a}$ in the
above limit, where $q_{2}$ is the gluon momentum, $T^b$ is the $SU(3)$ color matrix in the fundamental representation and $\Gamma^{a}$
represents the rest of the partonic scattering amplitude with
color indices for the two gluons $a$ and $b$. Similarly, Fig.~1(c)
can be reduced to: $-g/(-q_{2}^{+}+i\epsilon) \Gamma ^{a}T^{b}$.
The sum of these two diagrams will be ${g}/(-q_{2}^{+}+i\epsilon
)\left[T^b\Gamma^a-\Gamma^aT^b\right]$. Because of the unique
color index in $\Gamma _{a}$, we find the effective vertex as,
\begin{equation}
\mathrm{Fig.~1(b,c)}\sim \frac{i}{-q_{2}^{+}+i\epsilon
}(-ig)(-if_{bca})T^{c} \ ,
\end{equation}
which is exactly the first order expansion of the gauge link contribution in the gluon
distribution defined in Eq.~(\ref{g1}). Here, the crossing diagrams of Figs.1(b,c), by interchanging
the two gluons, are included in the calculations to compare to
the gauge link expansion results (see also Ref.~\cite{Collins:2008sg}). For all high order contributions, we can
follow the procedure outlined in Ref.~\cite{Belitsky:2002sm} to derive the
gluon distribution. 
Therefore, we obtain the following differential cross section
for the quark-antiquark jet correlation in DIS process
\begin{equation}
\frac{d\sigma^{\gamma_T^{\ast }A\rightarrow q\bar{q}+X}}{d{\cal P.S.}}
=\delta (x_{\gamma ^{\ast }}-1)
x_{g}G^{(1)}(x_{g},q_{\perp })H_{\gamma_T^*g\to q\bar q}\ ,\label{disdj}
\end{equation}%
where $x_g$ is the momentum fraction carried by the gluon and is
determined by the kinematics, $x_{\gamma^*}=z_q+z_{\bar q}$ with
$z_q$ and $z_{\bar q}$ being the momentum fractions of the virtual
photon carried by the quark and antiquark, respectively. The phase
space factor is defined as $d{\cal P.S.}= dy_{1}dy_{2}d^2P_{\perp
}d^{2}q_{\perp }$, and $y_1$ and $y_2$ are rapidities of the two
outgoing particles. The leading order hard partonic cross section
reads $H_{\gamma_T^*g\to q\bar q}={\alpha_{s}\alpha
_{em}e_{q}^{2}}(\hat s^2+Q^4)/(\hat s+Q^2)^4\times
\left({\hat{u}}/{\hat{t}}+{\hat{t}}/{\hat{u}}\right)$ with the
usually defined partonic Mandelstam variables $\hat s$, $\hat t$
and $\hat u$. By taking $Q^2=0$, we can extend the above result to the case of dijet productions in real photons scattering on nuclei. The above process (\ref{dis}) can also be analyzed
following the procedure in Ref.~\cite{Bomhof:2006dp}, where the
gluon distribution Eq.~(\ref{g1}) is written in the fundamental
representation,
\begin{eqnarray}
xG^{(1)}(x,k_\perp)&=&2\int \frac{d\xi ^{-}d\xi _{\perp }}{(2\pi )^{3}P^{+}}%
e^{ixP^{+}\xi ^{-}-ik_{\perp }\cdot \xi _{\perp }}\nonumber\\
&&\times \langle P|\textrm{Tr}\left[F^{+i}(\xi
^{-},\xi _{\perp })\mathcal{U}^{[+]\dagger }F^{+i}(0)\mathcal{U}%
^{[+]}\right]|P\rangle \ ,
\end{eqnarray}%
where the gauge link ${\cal
U}_\xi^{[+]}=U^n\left[0,+\infty;0\right]U^n\left[+\infty, \xi^{-};
\xi_{\perp}\right]$ with $U^n$ being the light-like Wilson line in
covariant gauge.

The quark-antiquark jet correlation in DIS (\ref{dis}) can also be calculated
directly in the color-dipole and CGC formalism~\cite{Gelis:2001da}. The basic element is
the $q\bar q$ (color-dipole) wave function of the
virtual photon, combined with the multi-scattering of the dipole on
the nuclear target. Following this formalism, the amplitude can be
written as
\begin{eqnarray}
|{\cal A}|^2&=&N_{c}\alpha _{em}e_{q}^{2}
\int \frac{\text{d}^{2}x}{(2\pi )^{2}}\frac{\text{d}^{2}x^{\prime }%
}{(2\pi )^{2}}\frac{\text{d}^{2}b}{(2\pi )^{2}}\frac{\text{d}^{2}b^{\prime }%
}{(2\pi )^{2}}e^{-ik_{1\perp }\cdot (x-x^{\prime })}\nonumber\\
&&\times e^{-ik_{2\perp }\cdot
(b-b^{\prime })}  \sum \psi_{T}^{\ast }(x-b)\psi_{T} (x^{\prime }-b^{\prime })\nonumber\\
&&\times \left[1+S^{(4)}_{x_g}(x,b;b^{\prime },x^{\prime
})-S^{(2)}_{x_g}(x,b)-S^{(2)}_{x_g}(b^{\prime },x^{\prime })\right]  \ ,
\end{eqnarray}%
where $\psi_T$ is the $q\bar q$ Fock state wave function of a
transversely polarized photon depending on
$\epsilon_f^2=z(1-z)Q^2$ with $z=z_q$,
$S_{x_g}^{(2)}(x,b)=\frac{1}{N_c}\left\langle\text{Tr}U(x)U^\dagger(b)\right\rangle_{x_g}$,
$S_{x_g}^{(4)}(x,b;b',x')=\frac{1}{N_c}\left\langle\text{Tr}U(x)
U^\dagger(x')U(b')U^\dagger(b)\right\rangle_{x_g}$, and $U$ is the
Wilson line describing the multiple scatterings of a single quark
with the nuclear target. The expectation value of multiple Wilson
lines can be found in
Refs.~\cite{Gelis:2001da,Dominguez:2008aa,{Baier:2005dv}}. In
order to study the differential cross section in the
correlation limit, we substitute $u=x-b$ and $v=zx+(1-z)b$. Thus,
the exponential factor becomes $e^{-iq_\perp\cdot
(v-v')}e^{-i\tilde P_\perp\cdot(u-u')}$ where $\tilde
P_\perp=(1-z)k_{1\perp}-zk_{2\perp}\approx P_\perp$. Then, we can
expand the interaction part of the bracket at small $u$ and $u'$,
but keep $v$ and $v'$ fixed. We find that the remaining
contribution comes from the term involving inelastic scatterings
while the elastic scattering part cancels out. Therefore, the
square bracket in the above equation becomes
$\frac{1}{N_c}u_iu'_j\left\langle\text{Tr}\partial_iU(v)U^\dagger(v')
\partial_jU(v')U^\dagger(v)\right\rangle_{x_g}$.
%\begin{equation}.
%\end{equation}
With this expansion result, we further find that the wave function
integral with $u_iu_j'$ leads to
$\delta_{ij} (P_\perp^4+\epsilon_f^4)/(P_\perp^2+\epsilon_f^2)^2$,
and the differential cross section can be simplified as,
\begin{eqnarray}
\frac{d\sigma ^{\gamma_T ^{\ast }A\rightarrow q\bar{q}X}}{d{\cal P.S.}}
&=&\alpha_{em}e_{q}^{2}\alpha_s\delta
\left(x_{\gamma^*}-1\right)z(1-z)\left(z^{2}+(1-z)^{2}\right)
\frac{P_\perp^4+\epsilon_f^4}{(P_{\perp }^{2}+\epsilon_f^2)^4}
\notag \\
&&\times
(16\pi^3)\int \frac{d^{3}v d^3v^{\prime }}{(2\pi )^{6}}e^{-iq_{\perp }\cdot (v-v^{\prime })}2\left\langle\text{Tr}F^{i+}({v})\mathcal{U}^{[+]\dagger}F^{i+}({v}')
\mathcal{U}^{[+]}\right\rangle_{x_g} \ .
\end{eqnarray}
%where the Wilson line $\mathcal{U}$ is the same as in Eq.~(7).
To compare with the differential cross section in Eq.~(\ref{disdj}),
we notice that the partonic Mandelstam variables can be
expressed in terms of $P_\perp$ and $z$ as:
$\hat s=P_\perp^2/(z(1-z))$, $\hat t=-(P_\perp^2+\epsilon_f^2)/(1-z)$,
and $\hat u=-(P_\perp^2+\epsilon_f^2)/z$.
Substituting these relations into Eq.~(9), taking the small-$x$
approximation for the gluon distribution, and correcting the normalization
for the states in the calculation of the associated matrix elements,
we find that it agrees with the factorization result (\ref{disdj})
completely. This consistency is very important to demonstrate
that we have a unified picture for the quark-antiquark correlation in DIS.

%\begin{figure}[tbp]
%\begin{center}
%\includegraphics[width=12cm]{qcd-photonjet}%\hfill
%\includegraphics[width=4cm]{glu3}%\hfill
%\end{center}
%\caption[*]{Photon-jet correlation in $pA$ collisions: (a) leading order,
%where the bubble represents the gluon attachments to both incoming and
%outgoing quark lines; (b,c) two-gluon exchange diagrams.}
%\label{fig2-1}
%\end{figure}

On the other hand, the direct photon-quark jet correlation in $pA$ collisions,
\begin{equation}
pA\rightarrow \gamma (k_{1})+q(k_{2})+X\ ,\label{dy}
\end{equation}%
probes a different gluon distribution.
We plot the relevant diagrams in Fig.~1 (d,e,f), again for the leading one gluon
exchange and two gluon exchanges. Similarly, the two gluon exchange
contributions can be summarized as
\begin{equation}
\mathrm{Fig.~1(e,f)}\sim \frac{i}{-q_{2}^{+}+i\epsilon }(-ig)\left(
T^{b}\Gamma ^{a}+\Gamma ^{a}T^{b}\right) \ ,\label{col2}
\end{equation}%
where the plus sign comes from the fact that the second gluon attaches to
the quark line in the initial and final states. Since there is no color
structure corresponding to Eq.~(\ref{col2}), we can only express it in the fundamental
representation. Following Ref.~\cite{Bomhof:2006dp}, we find that the gluon distribution in
this process can be written as
\begin{eqnarray}
xG^{(2)}(x,k_{\perp }) &=&2\int \frac{d\xi ^{-}d\xi _{\perp
}}{(2\pi )^{3}P^{+}}e^{ixP^{+}\xi ^{-}-ik_{\perp }\cdot \xi
_{\perp }}\langle
P|\textrm{Tr}\left[F^{+i}(\xi ^{-},\xi _{\perp })\mathcal{U}^{[-]\dagger }F^{+i}(0)\mathcal{U}%
^{[+]}\right]|P\rangle  \ ,\label{g2}
\end{eqnarray}
where the gauge link ${\cal U}_\xi^{[-]}=U^n\left[0,-\infty;0\right]U^n\left[-\infty, \xi^{-}; \xi_{\perp}\right]$.
This gluon distribution can also be calculated in the CGC formalism where it is found to be
$xG^{(2)}(x,q_\perp)\simeq \frac{q_{\perp }^{2}N_{c}}{2\pi^2 \alpha_s}S_{\perp
}F_{x_{g}}(q_{\perp }) $
with the normalized unintegrated gluon distribution
$F_{x_g}(q_\perp)=\int \frac{d^2r_\perp}{(2\pi)^2}e^{-iq_\perp\cdot r_\perp}
S_{x_g}^{(2)}(0,r_\perp)$.
Therefore, the differential cross section of (\ref{dy}) can be written as
\begin{equation}
\frac{d\sigma^{\left( pA\rightarrow \gamma q+X\right)} }{d{\cal P.S.}}
=\sum_{f}x_{1}q(x_{1})x_gG^{(2)}(x_g,q_{\perp })H_{qg\to\gamma q}\ ,
\end{equation}%
where $q(x_1)$ is the integrated quark distribution from the projectile
nucleon. Because we are taking large nuclear number limit, the
intrinsic transverse momentum associated with it can be neglected compared
to that from the gluon distribution of nucleus.
The hard partonic cross section is $H_{qg\to \gamma q}={\alpha_s\alpha_e e_q^2}/{(N_c\hat
s^2)}\left(-\hat s/\hat u-\hat u/\hat s\right)$.
We can calculate process (\ref{dy}) in the CGC formalism directly~\cite{Gelis:2002ki}, and again we find that these two calculations are consistent with each other.

The gluon distributions defined in Eqs.~(\ref{g1}) and (\ref{g2}) have the same
perturbative behavior and they reduce to the same gluon distribution after integrating over $k_{\perp}$. However, they do differ in the low transverse
momentum region~\cite{{Iancu:2003xm},Kharzeev:2003wz}.
It will be very important to measure the low $k_\perp$
behavior of the correlation processes of (\ref{dis}) and (\ref{dy}) to test these
predictions. In particular, the planed EIC machine
will be able to study the quark-antiquark jet correlation
in DIS process, whereas RHIC and future LHC experiments shall
provide information on direct photon-quark jet correlation in $pA$
collisions.

The extension to the dijet-correlation in $pA$ collisions is
straightforward.
The relevant initial and final state interaction phases
have been calculated in Ref.~\cite{Bomhof:2006dp}.
To obtain the differential cross section in the correlation
limit, we have to take the large $N_c$ limit and the mean field
approximation~\cite{Xiao:2010sp,{Fabio}}.
After a lengthy calculation, we find
\begin{eqnarray}
&&\frac{d\sigma^{(pA\rightarrow \mathrm{Dijet}+X)}}{d{\cal P.S.}}\nonumber\\
&&~~=\sum_{q}x_{1}q(x_{1})\frac{\alpha_s^2}{\hat{s}^{2}}\left[
\mathcal{F}_{qg}^{(1)}H_{qg\rightarrow qg}^{(1)}+\mathcal{F}_{qg}^{(2)}H_{qg\rightarrow qg}^{(2)}\right]   \notag \\
&&~~+x_{1}g(x_{1})\frac{\alpha_s^2}{\hat{s}^{2}}\left[
\mathcal{F}_{gg}^{(1)}\left( H_{gg\rightarrow
q\bar{q}}^{(1)}+\frac{1}{2} H_{gg\rightarrow
gg}^{(1)}\right) \right.   \notag \\
&&~~\left. +\mathcal{F}_{gg}^{(2)}\left( H_{gg\rightarrow q\bar{q}%
}^{(2)}+\frac{1}{2}H_{gg\rightarrow gg}^{(2)}\right) +\frac{1}{2}\mathcal{F}%
_{gg}^{(3)}H_{gg\rightarrow gg}^{(3)}\right] \ ,\label{dijet}
\end{eqnarray}
where again $q(x_1)$ and $g(x_1)$ are integrated quark and gluon
distributions from the projectile nucleon.
The hard partonic cross sections are defined as
\begin{eqnarray}
H_{qg\rightarrow qg}^{(1)}&=&
\frac{\hat{u}^{2}\left( \hat{s}^{2}+\hat{u}^{2}\right) }{-2\hat{s}\hat{u}\hat{t}^{2}},~~
H_{qg\rightarrow qg}^{(2)}=\frac{\hat{s}^{2}\left( \hat{s}^{2}+\hat{u}%
^{2}\right) }{-2\hat{s}\hat{u}\hat{t}^{2}} \nonumber\\
H_{gg\rightarrow q\bar{q}}^{(1)} &=&\frac{1}{4N_c}\frac{2\left( \hat{t}^{2}+%
\hat{u}^{2}\right) ^{2}}{\hat{s}^{2}\hat{u}\hat{t}},~~H_{gg\rightarrow q\bar{q}}^{(2)} =%
\frac{1}{4N_{c}}\frac{4\left( \hat{t}^{2}+\hat{u}^{2}\right) }{\hat{s}^{2}} \nonumber\\
H_{gg\rightarrow gg}^{(1)} &=&\frac{2\left( \hat{t}^{2}+\hat{u}^{2}\right) \left( \hat{s}^{2}-\hat{t}\hat{u}\right) ^{2}}{%
\hat{u}^{2}\hat{t}^{2}\hat{s}^{2}} , ~~
H_{gg\rightarrow gg}^{(2)} =\frac{4\left( \hat{s}^{2}-\hat{t}\hat{u}\right) ^{2}}{\hat{u}\hat{t}\hat{s}^{2}} \nonumber\\
H_{gg\rightarrow gg}^{(3)} &=&\frac{2\left( \hat{s}^{2}-\hat{t}\hat{u}\right) ^{2}}{\hat{u}^{2}\hat{t}^{2}},
\end{eqnarray}
and the various gluon distributions of nucleus $A$ are defined as
\begin{eqnarray}
\mathcal{F}_{qg}^{(1)} &=&xG^{(2)}(x,q_{\perp }),~~\mathcal{F}%
_{qg}^{(2)}=\int xG^{(1)}(q_1)\otimes F(q_2)\ ,  \notag \\
\mathcal{F}_{gg}^{(1)} &=&\int xG^{(2)}(q_1)\otimes F(q_2),~~\mathcal{F}%
_{gg}^{(2)}=-\int \frac{q_{1\perp}\cdot q_{2\perp}}{q_{1\perp}^2} xG^{(2)}(q_1)\otimes F(q_2)\ ,  \notag \\
\mathcal{F}_{gg}^{(3)} &=&\int xG^{(1)}(q_{1})\otimes F(q_2)\otimes F(q_3)\ ,
\end{eqnarray}%
where $\otimes $ represents the convolution in momentum space:
$\int \otimes=\int d^2q_{1}d^2q_2\delta^{(2)}(q_\perp-q_1-q_2)$.
Clearly, this process depends on both UGDs in a complicated way,
and the naive $k_t$-factorization does not hold.
We have checked the above results in different kinematics.
First, we recover the inclusive dijet cross section by integrating
over $q_\perp$. Second, in the dilute limit of $A$, or equivalently at
large $q_\perp$: $P_\perp \gg q_\perp \gg Q_S, \Lambda_{QCD}$, we reproduce
the dijet-correlation in the collinear factorization approach~\cite{qvy-short}.
Last but not least, we find the CGC calculation agrees with
above results perfectly.

Recently, both STAR and PHENIX Collaborations have published experimental results
on di-hadron correlations in $dAu$ collisions, where a strong back-to-back de-correlation
of the two hadrons was found in the forward rapidity region of the deuteron~\cite{rhic-data}.
These results have been compared to a number of theoretical calculations in the
CGC formalism, where different assumptions have been made
in the formulations~\cite{Marquet:2007vb, JalilianMarian:2004da, Tuchin:2009nf}. 
In particular, the numerical evaluation in Ref.~\cite{Marquet:2007vb}
only contains the first term in the $qg$ channel in Eq.(14). The
missing term is equally important and should be taken into
account to interpret the STAR data. We plan to compare our results to these calculations and those in
Ref.~\cite{{Blaizot:2004wv}} for $q\bar q$ production in $pA$ collisions
in a future publication\cite{Fabio}, together with detailed derivations of generalized results illustrated in this paper.

In summary, we have studied the $k_t$-factorization for dijet
production at small-$x$ in nuclei. We found that different gluon
distributions probed in different dijet production
processes can be built from two basic building blocks, the Weizs\"acker-Williams distribution and the Fourier transform of the dipole cross section. The most important result is that the DIS dijet process
can directly measure the well-known WW gluon distribution which
would be very interesting physics to pursue at the planned EIC.

We thank Al Mueller for stimulating discussions and critical
reading of the manuscript. We thank Larry McLerran, Jianwei Qiu,
and Raju Venugopalan for helpful conversations. We also thank
Cyrille Marquet for his collaborations at the early stage of this
work. This work was supported in part by the U.S. Department of
Energy under contracts DE-AC02-05CH11231. We are grateful to
RIKEN, Brookhaven National Laboratory and the U.S. Department of
Energy (contract number DE-AC02-98CH10886) for providing the
facilities essential for the completion of this work.


\begin{references}

%\cite{Boer:2003tx}
\bibitem{BoeVog03}
  D.~Boer and W.~Vogelsang,
  %``Asymmetric jet correlations in p p(pol.) scattering,''
  Phys.\ Rev.\ D {\bf 69}, 094025 (2004).
%  [arXiv:hep-ph/0312320].
  %%CITATION = HEP-PH 0312320;%%


\bibitem{Bomhof:2006dp}
  C.~J.~Bomhof, P.~J.~Mulders and F.~Pijlman,
  %``The construction of gauge-links in arbitrary hard processes,''
  Eur.\ Phys.\ J.\  C {\bf 47}, 147 (2006).
%  [arXiv:hep-ph/0601171].
  %%CITATION = EPHJA,C47,147;%%

\bibitem{qvy-short}
%\cite{Qiu:2007ar}
%\bibitem{Qiu:2007ar}
  J.~W.~Qiu, W.~Vogelsang and F.~Yuan,
  %``Asymmetric Di-jet Production in Polarized Hadronic Collisions,''
  Phys.\ Lett.\  B {\bf 650}, 373 (2007);
%  [arXiv:0704.1153 [hep-ph]].
  %%CITATION = PHLTA,B650,373;%%
%\cite{Qiu:2007ey}
%\bibitem{Qiu:2007ey}
%  J.~W.~Qiu, W.~Vogelsang and F.~Yuan,
  %``Single Transverse-Spin Asymmetry in Hadronic Dijet Production,''
  Phys.\ Rev.\  D {\bf 76}, 074029 (2007).
%  [arXiv:0706.1196 [hep-ph]].
  %%CITATION = PHRVA,D76,074029;%%


\bibitem{Collins:2007nk}
  J.~Collins and J.~W.~Qiu,
  %``$k_{T}$ factorization is violated in production of high-transverse-momentum
  %particles in hadron-hadron collisions,''
  Phys.\ Rev.\  D {\bf 75}, 114014 (2007).
%  [arXiv:0705.2141 [hep-ph]].
  %%CITATION = PHRVA,D75,114014;%%

%\cite{Vogelsang:2007jk}
\bibitem{Vogelsang:2007jk}
  W.~Vogelsang and F.~Yuan,
  %``Hadronic Dijet Imbalance and Transverse-Momentum Dependent Parton
  %Distributions,''
  Phys.\ Rev.\  D {\bf 76}, 094013 (2007);
%  [arXiv:0708.4398 [hep-ph]].
  %%CITATION = PHRVA,D76,094013;%%
%\bibitem{Collins:2007jp}
  J.~Collins,
  %``2-soft-gluon exchange and factorization breaking,''
  arXiv:0708.4410 [hep-ph].
  %%CITATION = ARXIV:0708.4410;%%

%\cite{Rogers:2010dm}
\bibitem{Rogers:2010dm}
  T.~C.~Rogers and P.~J.~Mulders,
  %``No Generalized TMD-Factorization in the Hadro-Production of High Transverse
  %Momentum Hadrons,''
  Phys.\ Rev.\  D {\bf 81}, 094006 (2010).
%  [arXiv:1001.2977 [hep-ph]].
  %%CITATION = PHRVA,D81,094006;%%

\bibitem{Xiao:2010sp}
  B.~W.~Xiao and F.~Yuan,
  %``Non-Universality of Transverse Momentum Dependent Parton Distributions at
  %Small-x,''
  Phys.\ Rev.\ Lett.\  {\bf 105}, 062001 (2010);
%  [arXiv:1003.0482 [hep-ph]].
  %%CITATION = PRLTA,105,062001;%%
%\bibitem{Xiao:2010sa}
%  B.~W.~Xiao and F.~Yuan,
  %``Initial and Final State Interaction Effects in Small-x Quark
  %Distributions,''
  arXiv:1008.4432 [hep-ph].
  %%CITATION = ARXIV:1008.4432;%%

%\cite{Gribov:1984tu}
\bibitem{Gribov:1984tu}
  L.~V.~Gribov, E.~M.~Levin and M.~G.~Ryskin,
  %``Semihard Processes In QCD,''
  Phys.\ Rept.\  {\bf 100}, 1 (1983).
  %%CITATION = PRPLC,100,1;%%

%\cite{Mueller:1985wy}
\bibitem{Mueller:1985wy}
  A.~H.~Mueller and J.~w.~Qiu,
  %``Gluon Recombination And Shadowing At Small Values Of X,''
  Nucl.\ Phys.\  B {\bf 268}, 427 (1986).
  %%CITATION = NUPHA,B268,427;%%

%\cite{McLerran:1993ni}
\bibitem{McLerran:1993ni}
  L.~D.~McLerran and R.~Venugopalan,
  %``Computing quark and gluon distribution functions for very large nuclei,''
  Phys.\ Rev.\  D {\bf 49}, 2233 (1994);
%  [arXiv:hep-ph/9309289].
  %%CITATION = PHRVA,D49,2233;%%
%\cite{McLerran:1993ka}
%\bibitem{McLerran:1993ka}
%  L.~D.~McLerran and R.~Venugopalan,
  %``Gluon distribution functions for very large nuclei at small transverse
  %momentum,''
  Phys.\ Rev.\  D {\bf 49}, 3352 (1994).
%  [arXiv:hep-ph/9311205].
  %%CITATION = PHRVA,D49,3352;%%

%\cite{Iancu:2003xm}
\bibitem{Iancu:2003xm}
E.~Iancu, A.~Leonidov and L.~McLerran,  %``The colour glass condensate: An introduction,''
 arXiv:hep-ph/0202270;
 E.~Iancu and R.~Venugopalan,
  %``The color glass condensate and high energy scattering in QCD,''
  arXiv:hep-ph/0303204;
  %%CITATION = HEP-PH/0303204;%%
%\cite{JalilianMarian:2005jf}
%\bibitem{JalilianMarian:2005jf}
  J.~Jalilian-Marian and Y.~V.~Kovchegov,
  %``Saturation physics and deuteron gold collisions at RHIC,''
  Prog.\ Part.\ Nucl.\ Phys.\  {\bf 56}, 104 (2006);
%  [arXiv:hep-ph/0505052].
  %%CITATION = PPNPD,56,104;%%
%\cite{Gelis:2010nm}
%\bibitem{Gelis:2010nm}
  F.~Gelis, E.~Iancu, J.~Jalilian-Marian and R.~Venugopalan,
  %``The Color Glass Condensate,''
  arXiv:1002.0333 [hep-ph];
  %%CITATION = ARXIV:1002.0333;%%
and references therein.

%\cite{Deshpande:2005wd}
\bibitem{Deshpande:2005wd}
  A.~Deshpande, R.~Milner, R.~Venugopalan and W.~Vogelsang,
  %``Study of the fundamental structure of matter with an electron ion
  %collider,''
  Ann.\ Rev.\ Nucl.\ Part.\ Sci.\  {\bf 55}, 165 (2005).
%  [arXiv:hep-ph/0506148].
  %%CITATION = ARNUA,55,165;%%


\bibitem{Kharzeev:2003wz} see e.g.,
  D.~Kharzeev, Y.~V.~Kovchegov and K.~Tuchin,
  %``Cronin effect and high-p(T) suppression in p A collisions,''
  Phys.\ Rev.\  D {\bf 68}, 094013 (2003).
%  [arXiv:hep-ph/0307037].
  %%CITATION = PHRVA,D68,094013;%%



%\cite{Collins:1981uw}
\bibitem{Collins:1981uw}
  J.~C.~Collins and D.~E.~Soper,
  %``Parton Distribution And Decay Functions,''
  Nucl.\ Phys.\  B {\bf 194}, 445 (1982).
  %%CITATION = NUPHA,B194,445;%%

%\cite{Ji:2005nu}
\bibitem{Ji:2005nu}
  X.~d.~Ji, J.~P.~Ma and F.~Yuan,
  %``Transverse-momentum-dependent gluon distributions and semi-inclusive
  %processes at hadron colliders,''
  JHEP {\bf 0507}, 020 (2005).
%  [arXiv:hep-ph/0503015].
  %%CITATION = JHEPA,0507,020;%%


%\cite{Belitsky:2002sm}
\bibitem{Belitsky:2002sm}
A.~V.~Belitsky, X.~Ji and F.~Yuan,
%``Final state interactions and gauge invariant parton
% distributions,''
Nucl.\ Phys.\ B {\bf 656}, 165 (2003).
%%CITATION = HEP-PH 0208038;%%

%\cite{Kovchegov:1998bi}
\bibitem{Kovchegov:1998bi}
  Y.~V.~Kovchegov and A.~H.~Mueller,
  %``Gluon production in current nucleus and nucleon nucleus collisions in  a
  %quasi-classical approximation,''
  Nucl.\ Phys.\  B {\bf 529}, 451 (1998).
%  [arXiv:hep-ph/9802440].
  %%CITATION = NUPHA,B529,451;%%


\bibitem{Brodsky:2002ue}
  S.~J.~Brodsky, P.~Hoyer, N.~Marchal, S.~Peigne and F.~Sannino,
  %``Structure functions are not parton probabilities,''
  Phys.\ Rev.\  D {\bf 65}, 114025 (2002).
%  [arXiv:hep-ph/0104291].
  %%CITATION = PHRVA,D65,114025;%%

\bibitem{Collins:2008sg}
  J.~C.~Collins and T.~C.~Rogers,
  %``The Gluon Distribution Function and Factorization in Feynman Gauge,''
  Phys.\ Rev.\  D {\bf 78}, 054012 (2008).
 % [arXiv:0805.1752 [hep-ph]].
  %%CITATION = PHRVA,D78,054012;%%

\bibitem{Gelis:2001da} F.~Gelis and A.~Peshier,
%``Probing colored glass via q anti-q photoproduction,''
Nucl.\ Phys.\ A \textbf{697}, 879 (2002).
% [arXiv:hep-ph/0107142].
%%CITATION = NUPHA,A697,879;%%

\bibitem{Dominguez:2008aa}
  F.~Dominguez, C.~Marquet and B.~Wu,
  %``On multiple scatterings of mesons in hot and cold QCD matter,''
  Nucl.\ Phys.\  A {\bf 823}, 99 (2009).
%  [arXiv:0812.3878 [nucl-th]].

\bibitem{Baier:2005dv}
  R.~Baier, A.~Kovner, M.~Nardi and U.~A.~Wiedemann,
  %``Particle correlations in saturated QCD matter,''
  Phys.\ Rev.\  D {\bf 72}, 094013 (2005).
%  [arXiv:hep-ph/0506126].
  %%CITATION = PHRVA,D72,094013;%%

%\cite{Gelis:2002ki}
\bibitem{Gelis:2002ki}
  F.~Gelis and J.~Jalilian-Marian,
  %``Photon production in high energy proton nucleus collisions,''
  Phys.\ Rev.\  D {\bf 66}, 014021 (2002).
%  [arXiv:hep-ph/0205037].
  %%CITATION = PHRVA,D66,014021;%%


\bibitem{Fabio}
F.~Dominguez, C.~Marquet,  B.W. ~Xiao, F.~Yuan, to be published.

\bibitem{rhic-data}
%\cite{Braidot:2010ig}
%\bibitem{Braidot:2010ig}
  E.~Braidot, for the STAR Collaboration,
  %``Two Particle Correlations at Forward Rapidity in STAR,''
  arXiv:1008.3989 [nucl-ex];
  %%CITATION = ARXIV:1008.3989;%%
B.~ Meredith, for the PHENIX Collaboration, to appear;



\bibitem{Marquet:2007vb}
  C.~Marquet,
  %``Forward inclusive dijet production and azimuthal correlations in pA
  %collisions,''
  Nucl.\ Phys.\  A {\bf 796}, 41 (2007);
%  [arXiv:0708.0231 [hep-ph]].
  %%CITATION = NUPHA,A796,41;%%
%\cite{Albacete:2010pg}
%\bibitem{Albacete:2010pg}
  J.~L.~Albacete and C.~Marquet,
  %``Azimuthal correlations of forward di-hadrons in d+Au collisions at RHIC in
  %the Color Glass Condensate,''
  Phys.\ Rev.\ Lett.\  {\bf 105}, 162301 (2010).%arXiv:1005.4065 [hep-ph].
  %%CITATION = ARXIV:1005.4065;%%


\bibitem{JalilianMarian:2004da}
  J.~Jalilian-Marian and Y.~V.~Kovchegov,
  %``Inclusive two-gluon and valence quark-gluon production in DIS and p A,''
  Phys.\ Rev.\  D {\bf 70}, 114017 (2004).
  %[Erratum-ibid.\  D {\bf 71}, 079901 (2005)].
  %[arXiv:hep-ph/0405266].
  %%CITATION = PHRVA,D70,114017;%%


\bibitem{Tuchin:2009nf}
  K.~Tuchin,
  %``Rapidity and centrality dependence of azimuthal correlations in
  %Deuteron-Gold collisions at RHIC,''
  Nucl.\ Phys.\  A {\bf 846}, 83 (2010).
%  [arXiv:0912.5479 [hep-ph]].
  %%CITATION = NUPHA,A846,83;%%

%\cite{Blaizot:2004wv}
\bibitem{Blaizot:2004wv}
  J.~P.~Blaizot, F.~Gelis and R.~Venugopalan,
  %``High energy p A collisions in the color glass condensate approach. II:
  %Quark production,''
  Nucl.\ Phys.\  A {\bf 743}, 57 (2004).
%  [arXiv:hep-ph/0402257].
  %%CITATION = NUPHA,A743,57;%%



\end{references}
\end{document}